\documentclass[aps,prl,twocolumn,showpacs,preprintnumbers,amsmath,amssymb,amsfonts,nofootinbib]{revtex4}

\newcommand{\bse}{\begin{subequations}}
\newcommand{\ese}{\end{subequations}}
\newcommand{\be}{\begin{equation}}
\newcommand{\ee}{\end{equation}}
\newcommand{\bea}{\begin{eqnarray}}
\newcommand{\eea}{\end{eqnarray}}
\DeclareMathOperator{\dslash}{\partial\!\!\! /}

\DeclareMathOperator{\xslash}{x\!\!\!\! /}
\begin{document}
\preprint{IC/2008/044\cr arXiv:0808.1677}
\title{3 Dimensional ${\cal N}=8$ Supersymmetric Field Theory Revisited}
\author{Mahdi Torabian}
\email[]{mahdi@ictp.it}\homepage{http://users.ictp.it/~mahdi}
\affiliation{International Centre for Theoretical Physics, Strada Costiera 11, I-34014, Trieste, Italy}
\date{August 12, 2008}
\begin{abstract}
Inspired by ideas regarding Hermitian $N\times N$ matrix fields obeying a non-associative algebra, 3-dimensional ${\cal N}=8$ SUSic field theories are proposed to on-shell represent subalgebras of $OSp(8|2)$ and $OSp(8|4)$ groups of SUSY transformations. They are theories of 8 scalar and 8 spinor fields with Yukawa, quartic and sextic self-interactions. The actions as their $R$-symmetry exhibit only $SO(7)$ or $SO(4)\times SO(4)$ subgroups of full $SO(8)$ automorphisms. It is argued that the number of degrees of freedom scale like $N^{3/2}$. There also exists an extra ${\cal S}_N$ permutation symmetry group.
\end{abstract}
\maketitle
\section{I. Introduction and Conclusion}
There are multilateral motivations to study maximally supersymmetric (SUSic) field theories in 3 dimensions (D) \cite{Schwarz,Seiberg}. The maximal supersymmetry (SUSY) in a theory (without general covariance) means 16 different SUSY transformations which can be promoted to 32 in scale invariant theories. In 3D it corresponds to ${\cal N}=8$ SUSY, the automorphism subgroup of which is $SO(8)$.

In this note algebras of $OSp(8|2)$ (Poincare SUSY) and $OSp(8|4)$ (conformal SUSY) groups of transformations with all possible algebraic extensions are studied in details. Then, inspired by ideas regarding Hermitian $N\times N$ matrix fields obeying a non-associative algebra and at the same time relaxing $R$-symmetry to $SO(7)$ and $SO(4)\times SO(4)$ subgroups of $SO(8)$ automorphisms, 3-dimensional ${\cal N}=8$ SUSic field theories are proposed to on-shell represent SUSY algebras. They are interacting field theory of 8 scalar and 8 spinor fields with Yukawa, quartic and sextic self-interactions. They explicitly realize algebraic extensions. They will also be promoted to conformal field theories once the dimensionful mass parameter is turned off.

Ever since M-theory has been hypothesized as a consistent theory for quantum gravity \cite{Witten}, M2-branes came into attention as its fundamental degrees of freedom \cite{M2}. It is believed that the low energy effective theory of a system of M2-branes is a 3 dimensional field theory with maximal SUSY. There are attempts to propose such a theory as Matter-Chern-Simons gauge theories \cite{Schwarz,BL,Gustavsson}. It is proposed that the conformal field theory presented here would describe dynamics of a stack of coincident M2-branes. Indeed the given actions, upon torus compactification (dimensional reduction to 1D), seem to be related to the DLCQ Hamiltonian of IIB string theory with $N$ units of light-cone momenta on flat or AdS$_5\times$S$^5$ backgrounds \cite{TGMT,DLCQ}. Thus, they might be linked to theory of system of $N$ coincident M2-branes \cite{Susskind}. Another pieces of evidence come from the fact that the number of degrees of freedom of matrix fields taking values in the non-associative algebra scales like $N^{3/2}$ and there exists an extra ${\cal S}_N(N\geq 4)$ discrete symmetry. It is also believed that the proposed theory with $SO(3,2)\times SO(7)$ symmetry structure is linked to $SO(7)$ invariant solution of 11D SUGRA \cite{deWit}.

\section{II. The $OSp(8|2)$ and $OSp(8|4)$ SUSY groups}
SUSY algebra is an extension of classical Lie algebras by adding generators in spinor representations \cite{HLS}. In 3D, they are spinors of the Lorentz group $SO(1,2) = SL(2,R)\approx USp(2)$. It admits three real valued Pauli matrices $\sigma_\mu$ and a real 2 dimensional representation, ${\bf 2_r}$.

The so-called ${\cal N}$-extension to the SUSY algebras is achieved by introducing a set of independent fermionic generators. Phenomenologically, in 3D the maximal SUSY is ${\cal N}=8$. The SUSY algebra is enlarged by automorphism group $SO(8)$ which is necessary by the closure condition. The fermionic generators are in its representation as well. It has 3 irreducible 8 dimensional representation $\bf 8_v$, $\bf 8_s$ and $\bf 8_c$ which are equivalent by triality. Thus, the fermionic generators, $Q_\alpha^{\dot A}$, of maximal SUSY algebra are in $({\bf 2_r,8_c})$ representation of $Spin(1,2)\times Spin(8)$ bosonic subgroup of SUSY group $OSp(8|2)$.

A non-trivial extension to ${\cal N}$-extended SUSY algebra comes thorough adding new elements to the anti-commutation relation. Minimally, anti-commutations of SUSY generators give bosonic generators of the algebra the SUSY is based on. It is possible to modify this by adding a number of bosonic elements which are allowed by the algebra. These addons are called algebraic ``extensions''. Extensions may  \cite{HLS} or may not \cite{Sezgin-1,Townsend} commute with the rest of the algebra. The necessary and sufficient condition for the consistency of the algebra is its closure, {\it i.e.} all the generators satisfy different Jacobi identities. It implies that in the case of SUSY algebras not based on Poincare algebra \cite{Kac,Nahm-1,Nahm-2}, adding bosonic extensions demands adding fermionic extensions as well \cite{Meessen,Peeters,extensions}.

The algebra of fermionic generators comes via anti-commutators $\big\{Q_\alpha^{\dot A},Q_\beta^{\dot B}\big\}$. By product of representations \bea {\bf 2_r\otimes 2_r} &=& {\bf 1^a\oplus 3^s},\\\label{8c-8c-product} {\bf 8_c\otimes 8_c} &=& {\bf 1^s\oplus 28^a\oplus 35_c^s},\eea one can see what appear on the right-hand side of the anti-commutation relation. The superscripts stand for symmetric and anti-symmetric representations. Having multiplied representations and projected out to symmetric representations, one finds \be\begin{split} &\big[({\bf 1^a\oplus 3^s})\otimes(\bf 1^s\oplus 28^a\oplus 35^s_c)\big]_{\rm symmetric} =\qquad\quad \cr &\qquad\qquad\qquad\qquad{\bf (3^s,1^s)\oplus(1^a,28^a)\oplus(3^s,35^s_c)}.\end{split}\ee Thus group theoretically, maximally extended SUSY algebra would be  \be\label{Poincare-extended-algebra}\big\{Q_\alpha^{\dot A},Q_\beta^{\dot B}\big\} \!=\! ({\cal C}\sigma^\mu)_{\alpha\beta}P_\mu{\bf 1}^{\dot A\dot B} + {\cal C}_{\alpha\beta}\Gamma^{IJ}_{\dot A\dot B}{\bf J}^{IJ} + ({\cal C}\sigma^\mu)_{\alpha\beta}{\cal Y}^{\dot A\dot B}_\mu.\ee $P_\mu$ is the generator of spacetime translations in $\bf (3^s,1^s)$ representation. ${\bf J}^{IJ}$, sitting in $\bf(1^a,28^a)$ representation, contains the generator for infinitesimal rotations in ``$R$-space'' ($I=1,2,\dots 8$). It satisfies \be [{\bf J}_{IJ},{\bf J}_{KL}] = i\left(\delta_{[JK}{\bf J}_{I]L} - \delta_{[JL}{\bf J}_{I]K}\right).\ee
They act on fermionic generators as
\be [{\bf J}^{IJ},Q_\alpha^{\dot A}] = \Gamma^{IK}_{\dot A\dot B}Q^{\dot A}_\alpha,\ee where $\Gamma^{IJ}$ is commutator of $SO(8)$ gamma matrices.
Furthermore, ${\cal Y}_\mu^{\dot A\dot B}$ is a symmetric tensor in $\bf(3^s,35^s)$ representation. Note that ${\cal C}\sigma^\mu$ and is symmetric matrix and ${\cal C}$ is an anti-symmetric one. The matrix ${\cal C}$ is charge conjugation matrix, {\it i.e.} ${\cal C}\sigma_\mu{\cal C}^{-1}=-\sigma^t_\mu$. It worths noting that both sides of \eqref{Poincare-extended-algebra} has 136 independent component. For completeness, it is good to mention \be [M^{\mu\nu},Q_\alpha^{\dot A}] = \epsilon^{\mu\nu\rho}\sigma^\rho_{\alpha\beta}Q^{\dot A}_\beta,\ee where $M_{\mu\nu}$ is the generator of Lorentz transformations.

Consistency of the extended algebra demands its closure, {\it i.e.} satisfy Jacobi identities. In general extensions carrying spacetime and $R$-space indices would have non-trivial commutators with corresponding rotations generators (hence obviously they are not central to the original SUSY algebra). Therefor, closure of the algebra may imply that commutators is non-zero and forces to introduce new fermionic generators.
\paragraph{$SO(7)$ Decomposition} For latter use, one considers decompositions of representations of $SO(8)$ into representations of $SO(7)$ subgroup. \bea \bf 8_v&\longrightarrow& \bf 1+7,\\ \bf 8_{s,c}&\longrightarrow& \bf 8_r,\eea where $\bf 1, 7$ and $\bf 8_r$ are singlet, fundamental (vector) and spinor representations of $SO(7)$ respectively. Then \be \bf 8_r\otimes 8_r = 1^s\oplus 7^a\oplus 21^a\oplus 35^s.\ee Branching rules for other representations is as follows \be \bf 28^a \longrightarrow \bf 7+21 \qquad,\qquad \bf 35_{s,c} \longrightarrow \bf 35.\ee In this case, one reads \be\begin{split} &\big[({\bf 1^a\oplus 3^s})\otimes(\bf 1^s\oplus 7\oplus 21^a\oplus 35^s)\big]_{\rm symmetric} =\qquad\qquad \cr &\qquad\qquad{\bf (3^s,1^s)\oplus(1^a,7^a)\oplus(1^a,21^a)\oplus(3^s,35^s)}.\end{split}\ee
Thus, the algebra \eqref{Poincare-extended-algebra} is rewritten as \be \begin{split} \big\{Q_\alpha^A,Q_\beta^B\big\} &= ({\cal C}\sigma^\mu)_{\alpha\beta}P_\mu{\bf 1}^{AB} + {\cal C}_{\alpha\beta}\gamma^{ij}_{AB}{\bf J}^{ij} \cr &+ {\cal C}_{\alpha\beta}\gamma^i_{AB}{\cal Z}^i + ({\cal C}\gamma^\mu)_{\alpha\beta}{\cal Y}^{AB}_\mu, \end{split}\ee
where $i=1,\dots,7$ and $\gamma^i$ are gamma matrices of $SO(7)$.

\paragraph{Conformal Extension} The Poincare group can be promoted to conformal group by adding dilatation and special conformal transformations (SCT). In 3D it is locally isomorphic to group $SO(2,3)$. The generators of its algebra is denoted by $M_{mn}$ where $m$ runs from $-1$ to $3$ and satisfy the following algebra
\be [M_{mn},M_{pq}] = i\left(\eta_{[np}M_{m]q} - \eta_{[nq}M_{m]p}\right),\ee where $\eta_{mn}$ = {\sl diag}$(+---+)$. The SUSic conformal algebra is achieved by adding fermionic generators in conformal spinor representation, $\bf 4_r$. In fact a conformal spinor is composed of two real Lorentz spinors $\bf 2_r$, {\it i.e.} two $USp(2)$ spinors $Q_\alpha$ and $S_\alpha$ are completed to a $USp(4)$ spinor ${\cal Q}_a$. It satisfies \be [M_{mn},{\cal Q}_a] = \gamma^{mn}_{ab}{\cal Q}_b,\ee where $\gamma^{mn}=1/2[\gamma^m,\gamma^n]$ and $\gamma^m$ are gamma matrices of $SO(2,3)$. The number of fermionic generators of maximal conformal SUSY algebra is 32 which sit in $({\bf 4_r,8_c})$ representation of $USp(4)\times Spin(8)$ as bosonic subgroup of $OSp(8|4)$, and denoted by ${\cal Q}_a^{\dot A}$. It is proved that there can exist conformal supersymmetric field theories with this number of supercharges in 3D \cite{Nahm-1}.

The algebra of fermionic generators comes via anti-commutators $\big\{{\cal Q}_a^{\dot A},{\cal Q}_b^{\dot B}\big\}$. By product of representations \be {\bf 4_r\otimes 4_r} = {\bf 6^a\oplus 10^s},\ee one can see what may appear on the right-hand side of the anti-commutation relations. Having multiplied representations and projected to symmetric representations, one finds \be\begin{split} \big[({\bf 6^a\oplus 10^s})&\otimes(\bf 1^s\oplus 28^a\oplus 35^s_c)\big]_{\rm symmetric} = \cr &\qquad{\bf (10^s,1^s)\oplus(6^a,28^a)\oplus(10^s,35^s_c)}.\end{split}\ee
Thus, maximally extended SUSY algebra would be \be\begin{split} \big\{{\cal Q}_a^{\dot A},{\cal Q}_b^{\dot B}\big\} &= ({\cal C}\gamma^{mn})_{ab}M_{mn}{\bf 1}_{\dot A\dot B} + {\cal  C}_{ab}\Gamma^{IJ}_{\dot A\dot B}{\bf J}^{IJ} \cr &+ ({\cal C}\gamma^m)_{ab}\Gamma^{IJ}_{\dot A\dot B}{\cal Z}^{IJ}_m + ({\cal C}\gamma^{mn})_{ab}{\cal Y}^{\dot A\dot B}_{mn}. \end{split}\ee The extension ${\cal Y}_{mn}^{\dot A\dot B}$ is symmetric and anti-symmetric with respect to upper and lower indices respectively and is in representation $\bf(10^s,35_c^s)$. The extension ${\cal Z}^{IJ}_m$ in representation $\bf(6^a,28^a)$ is anti-symmetric with respect to upper indices. Note that ${\cal C}\gamma^{mn}$ is a symmetric matrix, ${\cal C}$ and ${\cal C}\gamma^m$ are anti-symmetric matrices. The matrix ${\cal C}$ is charge conjugation matrix in spacetime, {\it i.e.} ${\cal C}\gamma_m{\cal C}^{-1}=-\gamma^t_m$. It is interesting to note that both sides of the above relation has 528 independent component.

Again, consistency demands closure of the algebra. Due to existence of tensor extensions (which behave non-trivially under rotations generators $M$ and $\bf J$), the Jacobi identities demand some fermionic extensions.

\paragraph{Poincare Decomposition}
It is customary to decompose the conformal algebra into its Lorentz and dilation subgroup. This is due to the fact that only this subgroups are manifest in the level of the action of the field theory. The generator of rotations is partitioned as \be M_{mn}\rightarrow M_{-13},M_{-1\mu},M_{3\mu},M_{\mu\nu},\ee
where $M_{-13}$ is identified as $D$, the dilatation generator, and $M_{-1\mu}\pm M_{3\mu}$ is defined $P_\mu$ and $K_\mu$, generators of translations and SCT's respectively. Further, $M_{\mu\nu}$ is generator of rotations in 3 dimensional Minkowski spacetime.

The vector extension is decomposed as \be {\cal Z}_m^{IJ}\rightarrow{\cal Z}_{-1}^{IJ},{\cal Z}_3^{IJ},{\cal Z}_\mu^{IJ},\ee for later use, one defines
${\cal Z}^{IJ}_{-1}\pm{\cal Z}^{IJ}_3={\cal Z}^{IJ}_\pm$. The tensor extension is partitioned as \be {\cal Y}_{mn}^{\dot A\dot B}\rightarrow{\cal Y}_{-1\mu}^{\dot A\dot B}, {\cal Y}_{3\mu}^{\dot A\dot B}, {\cal Y}_{-13}^{\dot A\dot B}, {\cal Y}_{\mu\nu}^{\dot A\dot B}.\ee One define
${\cal Y}_{-13}^{\dot A\dot B}={\cal Y}^{\dot A\dot B}$ and combination ${\cal Y}_{-1\mu}^{\dot A\dot B}\pm{\cal Y}_{3\mu}^{\dot A\dot B}={\cal Y}_{\mu\pm}^{\dot A\dot B}$ for later convenience.

The fermionic generator ${\cal Q}_a^{\dot A}$ is also decomposed into two sets of generators, $Q_\alpha^{\dot A}$ and $S_\alpha^{\dot A}$ in representation $\bf (2_r,8_c)$. For ``$Q-Q$'' and ``$S-S$'' anti-commutators, upon imposing symmetric condition one finds \be \big\{Q_\alpha^{\dot A},Q_\beta^{\dot B}\big\} \!\!= ({\cal C}\sigma^\mu)_{\alpha\beta}P_\mu{\bf 1}^{\dot A\dot B} \!+  {\cal C}_{\alpha\beta}\Gamma^{IJ}_{\dot A\dot B}{\cal Z}^{IJ}_+ \!+ ({\cal C}\sigma^\mu)_{\alpha\beta}{\cal Y}^{\dot A\dot B}_{\mu+}, \ee \be \big\{S_\alpha^{\dot A},S_\beta^{\dot B}\big\} \!\!= ({\cal
C}\sigma^\mu)_{\alpha\beta}K_\mu{\bf 1}_{\dot A\dot B} + {\cal C}_{\alpha\beta}\Gamma^{IJ}_{\dot A\dot B}{\cal Z}^{IJ}_- \!+ ({\cal C}\sigma^\mu)_{\alpha\beta}{\cal Y}^{\dot A\dot B}_{\mu-}.\ee
Regarding ``$Q-S$'' anti-commutation relation, whereas there is no symmetry restriction on the representation, one finds that product of representations gives \be\label{Q-S-algebra}\begin{split} \big[\bf
(1^a\oplus3^s)\otimes(1^s\oplus28^a\oplus35^s_c)\big] = \bf(1^{\bf a},1^{\bf s})&\oplus\bf(1^{\bf a},28^{\bf a}) \cr \oplus\bf(1^{\bf a},35_c^s) \oplus \bf (3^{\bf s},1^{\bf s})\oplus\bf(3^{\bf s},28^{\bf a})&\oplus\bf(3^{\bf s},35_c^s).\end{split}\ee Explicitly, in terms of bosonic generators and extension one finds (noting $M_\mu=\epsilon_{\mu\nu\rho}M_{\nu\rho}$ and ${\cal Y}^{\dot A\dot B}_\mu=\epsilon_{\mu\nu\rho}{\cal Y}_{\mu\nu}^{\dot A\dot B}$)
\be\begin{split}\big\{Q_\alpha^{\dot A},S_\beta^{\dot B}\big\} &=
{\cal C}_{\alpha\beta}D{\bf 1}_{\dot A\dot B} + ({\cal
C}\sigma^\mu)_{\alpha\beta}M_\mu{\bf 1}_{\dot A\dot B} + {\cal C}_{\alpha\beta}
\Gamma^{IJ}_{\dot A\dot B}{\bf J}^{IJ} \cr &+ ({\cal C}\sigma^\mu)_{\alpha\beta}\Gamma^{IJ}_{\dot A\dot B}{\cal Z}^{IJ}_\mu + {\cal C}_{\alpha\beta}{\cal Y}^{\dot A\dot B} + ({\cal C}\sigma^\mu)_{\alpha\beta}{\cal Y}^{\dot A\dot B}_\mu.\end{split}\ee

\section{III. The Field Theory Realization}
Having studied ${\cal N}=8$ SUSY algebras in details, field theories are presented here to on-shell realize them. Naively, in order to fields theoretically represent the algebras, 8 real scalar fields and 8 spinor fields are needed. It is argued bellow that interacting actions can be written through two assumptions; dynamical fields are Hermitian matrices valued in a non-associative algebra and furthermore, $SO(8)$ automorphism is relaxed to its subgroups. Two cases are considered here; theories realizing $SO(7)$ and $SO(4)\times SO(4)$ groups as their global $R$-symmetry.

\subsubsection{$SO(7)$ subgroup of $SO(8)$}
As for bosonic field, one considers 8 scalar fields $\phi^i$ and $\phi^8$ in $\bf 7$ and $\bf 1$ representations of $SO(7)$ respectively. For fermionic counterparts, one includes 8 Majorana spinor fields $\psi^A_\alpha$ in  $\bf(2_r,8_r)$ of $Spin(1,2)\times Spin(7)$. The crucial point in using $SO(7)$ subgroup is that in addition the metric $\delta^{ij}$ and pseudo-tensor $\epsilon^{ijklmnp}$, $SO(7)$ group has another invariant anti-symmetric tensor $c^{ijk}$ (and its dual $c^{ijkl}=\epsilon^{ijklmnp}c_{mnp}/3!$) which can be used to propose non-linear terms. Furthermore, dynamical fields are supposed to be promoted to Hermitian matrix fields satisfying non-associative algebra of generalized gamma matrices of $Spin(4)$ [see appendix B]. This algebraic structure is implemented via generalized 4-commutators.

All things considered, the most general objects sitting in fundamental and singlet representation of $SO(7)$ and spinorial of $Spin(7)$ are \bea {\bf 7}&:& m\phi^i +gt^{ijkl}[\phi^j,\phi^k,\phi^l,{\cal T}]+gt^{ijk}[\phi^j,\phi^k,\phi^8,{\cal T}],\cr {\bf 1}&:&m\phi^8+gt^{jkl}[\phi^j,\phi^k,\phi^l,{\cal T}],\cr {\bf 8_r}&:&m\psi^A + g\gamma^{ij}_{AB}[\phi^i,\phi^j,\psi^B,{\cal T}] + g\gamma^i_{AB}[\phi^i,\phi^8,\psi^B,{\cal T}],\nonumber\eea
where ${\cal T}$ is a fixed matrix and 4-commutator is just fully anti-symmetric matrix product of 4 matrices\cite{TGMT,half-BPS},[appendix B]. $\gamma^i$ are $Spin(7)$ gamma matrices satisfying Dirac algebra $\{\gamma^i,\gamma^j\}=2\delta^{ij}$ and $\gamma^{ij}=1/2[\gamma^i,\gamma^j]$. Later on, constants $m$ and $g$ will be regarded as mass and coupling constant.

One proposes the following as the most general renormalizable action governing the dynamics of fields \be\label{massive-action}\begin{split} \!\!\!{\cal S}
=& \int d^3x{\rm Tr}\bigg[\frac{1}{2}\partial^\mu\phi^i\partial_\mu\phi^i + \frac{1}{2}\partial^\mu\phi^8\partial_\mu\phi^8 +
i\bar\psi^A\sigma^\mu\partial_\mu\psi^A \cr &- \frac{1}{2}\Big(m\phi^i+gt^{ijkl}[\phi^j,\phi^k,\phi^l,{\cal T}]+gt^{ijk}[\phi^j,\phi^k,\phi^8,{\cal T}]\Big)^2 \cr &- \frac{1}{2}\Big(m\phi^8+gt^{jkl}[\phi^j,\phi^k,\phi^l,{\cal T}]\Big)^2 \cr -& \bar\psi^A\!\Big(m\psi^A\! + g\gamma^{ij}_{AB}[\phi^i,\phi^j,\psi^B,{\cal T}]\! + g\gamma^i_{AB}[\phi^i,\phi^8,\psi^B,{\cal T}]\Big).\end{split}\ee It is manifestly invariant under $ISO(1,2)\times SO(7)$ group of transformations, as well as discrete ${\cal S}_N$ group of permutations of $N$ objects (see [appendix A] for explicit construction of the action).  Note that to avoid confusion, some numerical coefficients (1/3! and 1/2) are ignored here, they can be easily reimbursed by noting product of anti-symmetric tensors and 4-commutators.

The equations of motion can be derived as
\be\begin{split} \square\phi^i = &-m^2\phi^i \cr & - g\gamma^{ij}_{AB}[\bar\psi^A,\psi^B,\phi^j,{\cal T}] - g\gamma^i_{AB}[\bar\psi^A,\psi^B,\phi^8,{\cal T}] \cr &- mgt^{ijkl}[\phi^j,\phi^k,\phi^l,{\cal T}] - mgt^{ijk}[\phi^j,\phi^k,\phi^8,{\cal T}] \cr - g^2[\phi^j&,\phi^k,[\phi^i,\phi^j,\phi^k,{\cal T}],{\cal T}]\! -\! g^2[\phi^j,\phi^8,[\phi^i,\phi^j,\phi^8,{\cal T}],{\cal T}], \end{split}\ee
\vspace*{-6mm}\be\begin{split} \square\phi^8 = &-m^2\phi^8 \\ & - g\gamma^i_{AB}[\bar\psi^A,\psi^B,\phi^i,{\cal T}] - mgt^{ijk}[\phi^i,\phi^j,\phi^k,{\cal T}] \cr &- g^2[\phi^j,\phi^k,[\phi^8,\phi^j,\phi^k,{\cal T}],{\cal T}], \end{split}\ee
\vspace*{-6mm}\be\begin{split} \sigma^\mu\partial_\mu\psi^A = &-m\psi^A \cr &- g\gamma^{ij}_{AB}[\phi^i,\phi^j,\psi^B,{\cal T}] - g\gamma^i_{AB}[\phi^i,\phi^8,\psi^B,{\cal T}].\end{split} \ee

One discovers the following set of transformations relating bosonic and fermionic degrees of freedom, thus called SUSY transformations, that leave action intact
\bea\label{bosonic-Poincare-SUSY} \delta\phi^i &=& \bar\epsilon\gamma^i\psi,\\ \delta\phi^8 &=& \bar\epsilon\psi,\eea
\be\begin{split} \delta\psi = \big(\dslash\phi^i + m\phi^i &+ gt^{ijk}[\phi^j,\phi^k,\phi^8,{\cal T}] \cr &+ gt^{ijkl}[\phi^j,\phi^k,\phi^l,{\cal T}]\big)\gamma^i\epsilon \cr + \big(\dslash\phi^8 + m\phi^8 &+ gt^{ijk}[\phi^i,\phi^j,\phi^k,{\cal T}]\big)\epsilon. \end{split}\ee
However, invariance of the actin is necessary but not enough. Using equations of motion, one must check the on-shell closure of SUSY transformations on fields. Applying twice on fields and forming commutators one finds
\be\begin{split} &[\delta_\epsilon,\delta_{\epsilon'}]\phi^i = \iota^\mu\partial_\mu\phi^i \cr &+ \Omega_{ij}(m\phi^j+gt^{jklm}[\phi^k,\phi^l,\phi^m,{\cal T}] + t^{jkl}[\phi^k,\phi^l,\phi^8,{\cal T}])
\cr &+ \Omega_i(m\phi^8+gt^{jkl}[\phi^j,\phi^k,\phi^l,{\cal T}])
,\end{split}\ee
\vspace*{-4mm}\be\begin{split} &[\delta_\epsilon,\delta_{\epsilon'}]\phi^8 = \iota^\mu\partial_\mu\phi^8
\cr &+ \Omega_i(m\phi^i+gt^{ijkl}[\phi^j,\phi^k,\phi^l,{\cal T}] + t^{ijk}[\phi^j,\phi^k,\phi^8,{\cal T}])
,\end{split}\ee
\vspace*{-5mm}\be\label{fermionic-sector}\begin{split} &[\delta_\epsilon,\delta_{\epsilon'}]\psi^A = \iota^\mu\partial_\mu\psi^A +
(\Omega_{ij}\gamma^{ij}_{AB} + \Omega_i\gamma^i_{AB})
\times\cr &\ \times(m\psi^B+g\gamma^{kl}_{BC}[\phi^k,\phi^l,\psi^C,{\cal T}] + g\gamma^k_{BC}[\phi^k,\phi^8,\psi^C,{\cal T}]),\end{split}\ee where $\iota^\mu$ and $\Omega_{ij}$ are parameters of translation and $SO(7)$ rotations. $\Omega_i$ is an extra parameter, showing up because of bosonic mixing in fermionic sector \eqref{fermionic-sector}. One sees they close on translation and rotations. They also imply some algebraic extensions which becomes more clear soon.

The SUSY conserved charge can be written as
\be\begin{split} Q^A\!\!=\!\!\int\!{\rm d}^2x
{\rm Tr}\Big[\!\Big(\!\big(\!\dslash\phi^i\! -i m\phi^i &\!\!-\!igt^{ijkl}[\phi^j,\phi^k,\phi^l,{\cal T}] \cr &\!\!-\!igt^{ijk}[\phi^j,\phi^k,\phi^8,{\cal T}]\big)\gamma^i\cr + ( \dslash\phi^8 -i\! m\phi^8 &\!\!-\!i gt^{ijk}[\phi^i,\phi^j,\phi^k,{\cal T}])\Big)({\cal C}\psi^A)\Big].\end{split}\ee
Using canonical commutation relations of canonical fields and their conjugate momenta \bea [(\phi^i)_{pq},(\pi^j)_{rs}] &=&
i\delta^{ij}\ \delta_{ps}\delta_{qr}, \\\ [(\phi^8)_{pq},(\pi^8)_{rs}] &=&
i\delta_{ps}\delta_{qr}, \\
\big\{(\psi^A_\alpha)_{pq},(\bar\psi^B_\beta)_{rs}\big\} &=& \delta^{AB}
\delta_{\alpha\beta}\delta_{ps}\delta_{qr},\eea one can evaluate the ``$Q-Q$'' anti-commutator of SUSY charges and read other conserved charges. 

The Hamiltonian and physical momentum can be read as follows \be\begin{split} \!\!\!H =& \int{\rm d}^2x{\rm Tr}\Big[\frac{1}{2}\pi^i\pi^i + \frac{1}{2}\pi^i\pi^i + \frac{1}{2}\partial_i\phi^i\partial_i\phi^i + \frac{1}{2}\partial_i\phi^i\partial_i\phi^i \cr -& i\psi^{\dagger A}\sigma^i\partial_i\psi^A  +  \frac{1}{2} m^2\phi^i\phi^i + \frac{1}{2} m^2\phi^8\phi^8 + m\bar\psi^A\psi^A \cr +&  g\gamma^{ij}_{AB} \bar\psi^A[\phi^i,\phi^j,\psi^B,{\cal T}] + g\gamma^i_{AB} \bar\psi^A[\phi^i,\phi^8,\psi^B,{\cal T}] \cr + & mgt^{ijkl}\phi^i[\phi^j,\phi^k,\phi^l,{\cal T}] + mgt^{ijk}\phi^i[\phi^j,\phi^k,\phi^8,{\cal T}] \cr +& g^2[\phi^i,\phi^j,\phi^k,{\cal T}]^2 + g^2[\phi^i,\phi^j,\phi^8,{\cal T}]^2 \Big],\end{split}\ee
\vspace*{-7mm}\be {\rm P}_m = \int {\rm d}^2x {\rm Tr}\Big[\pi^i\partial_m\phi^i + \pi^8\partial_m\phi^8 + \psi^{\dagger A}\partial_m\psi^A\Big],\ee where $\pi^{i,8}=\partial_0\phi^{i,8}$ are field conjugate momenta. They together constituent 3-momentum $P_\mu=(H,{\rm P}_m)$ sitting in representation $\bf(3^s,1^s)$. One derives the conserved charge of rotations in $R$-space in representation $\bf(1^a,21^a)$ and an algebraic extension in $\bf(1^a,7^a)$ as \bea J^{ij} &=& m\int d^2x{\rm
Tr}\Big[\pi^{[i}\phi^{j]} + i\bar\psi^A\gamma^{ij}_{AB}\psi^B\Big],\\ J^i_\pm &=& \pm\ m\int d^2x{\rm Tr}\Big[\pi^{[i}\phi^{8]} + i\bar\psi^A\gamma^i_{AB}\psi^B\Big].\eea

There exists other algebraic extensions in representation  $\bf(1^a,21^a)$ and $\bf(1^a,7^a)$ which consist of two parts. One is due to matrix nature of canonical fields as \be\begin{split} {\cal Z}^{ij}_{\rm matrix} = \int {\rm d}^2x{\rm
Tr}\Big[\big(\pi^{[i}+m\phi^{[i}\big)& \cr \times g\!\big(t^{j]klm}[\phi^k,\phi^l,\phi^m,{\cal T}] &+ t^{j]kl}[\phi^k,\phi^l,\phi^8,{\cal T}]\big)\cr + g\bar\psi^A\gamma^{ij}_{AB}\big(\gamma^{kl}_{BC}[\phi^k,\phi^l,\psi^C,{\cal T}] &+ \gamma^k_{BC}[\phi^k,\phi^8,\psi^C,{\cal T}]\big)\Big], \end{split}\ee
\be\begin{split} {\cal Z}^i_{\pm\rm matrix} = \pm\int d^2x{\rm
Tr}\Big[\big(\pi^{[i}+m\phi^{[i}\big) gt^{8]jkl}&[\phi^j,\phi^k,\phi^l,{\cal T}] \cr + g\bar\psi^A\gamma^i_{AB}\big(\gamma^{kl}_{BC}[\phi^k,\phi^l,\psi^C,{\cal T}] + \gamma^k_{BC}&[\phi^k,\phi^8,\psi^C,{\cal T}]\big)\Big]. \end{split}\ee
The other is a topological extension which is non-vanishing only for a configuration with non-trivial boundary conditions
\bea {\cal Z}^{ij}_{\rm boundary} &=& \int {\rm d}^2x\epsilon^{0mn}{\rm
Tr}\Big[\partial_m\phi^i\partial_n\phi^j\Big],\\ {\cal Z}^i_{\rm boundary} &=& \int {\rm d}^2x\epsilon^{0mn}{\rm
Tr}\Big[\partial_m\phi^i\partial_n\phi^8\Big].\eea

Furthermore, there exists algebraic extension in $\bf(3^s,35^s_c)$ representation as follows
\be\begin{split}{\cal Y}^{\dot A\dot B}_{\mu} = \int {\rm d}^2x g\epsilon^{\mu 0m}\partial_m{\rm Tr}\Big[&\phi^it^{ijkl}[\phi^j,\phi^k,\phi^l,{\cal T}]\cr +&\phi^8t^{ijk}[\phi^i,\phi^j,\phi^k,{\cal T}]\Big]{\bf 1}^{\dot A\dot B},\end{split}\ee the above extension is there because of both the matrix nature of canonical fields and non-trivial boundary.

\paragraph{Scale Invariant Theory}
By turning off the dimensionful parameter in the action \eqref{massive-action}, {\it i.e.} the mass $m$, the action becomes scale invariant. It is believed that the in a unitary interacting theory, the Hilbert space will be symmetric under full conformal group of transformations. The action of the conformal field theory would be \be\label{confornal-action}\begin{split} \!\!\!{\cal S}
=& \int d^3x{\rm Tr}\Big[\frac{1}{2}\partial^\mu\phi^i\partial_\mu\phi^i + \frac{1}{2}\partial^\mu\phi^8\partial_\mu\phi^8 +
i\bar\psi^{\dot A}\sigma^\mu\partial_\mu\psi^{\dot A} \cr &- \frac{1}{2}\Big(gt^{ijkl}[\phi^j,\phi^k,\phi^l,{\cal T}]+gt^{ijk}[\phi^j,\phi^k,\phi^8,{\cal T}]\Big)^2 \cr &- \frac{1}{2}\Big(gt^{jkl}[\phi^j,\phi^k,\phi^l,{\cal T}]\Big)^2 \cr &- \bar\psi^A\Big(g\gamma^{ij}_{AB}[\phi^i,\phi^j,\psi^B,{\cal T}] + g\gamma^i_{AB}[\phi^i,\phi^8,\psi^B,{\cal T}]\Big)\Big].\end{split}\ee

There exists a set of Poincare SUSY transformations, which leave the action invariant. Bosonic transformations are as before \eqref{bosonic-Poincare-SUSY} but the fermionic one is
\be\begin{split} \delta\psi = &\big(\dslash\phi^i + gt^{ijk}[\phi^j,\phi^k,\phi^8,{\cal T}] + gt^{ijkl}[\phi^j,\phi^k,\phi^l,{\cal T}]\big)\gamma^i\epsilon \cr + &\big(\dslash\phi^8 + gt^{ijk}[\phi^i,\phi^j,\phi^k,{\cal T}]\big)\epsilon. \end{split}\ee
There exist also a set of conformal SUSY transformations that leave the action intact
\bea \delta_\varepsilon\phi^i &=&
\bar\varepsilon \xslash\gamma^i\psi,\\ \delta_\varepsilon\phi^8 &=&
\bar\varepsilon \xslash\gamma^8\psi,\eea
\be\begin{split}\delta_\varepsilon\psi =& -\big(\xslash\dslash\phi^i
+ \phi^i + \xslash t^{ijkl}[\phi^j,\phi^k,\phi^l,{\cal T}]\cr &\qquad\qquad\qquad\ + \xslash t^{ijk}[\phi^j,\phi^k,\phi^8,{\cal T}]\big)\gamma^i\varepsilon \cr & -\big(\xslash\dslash\phi^8
+ \phi^8 + \xslash t^{8jkl}[\phi^j,\phi^k,\phi^l,{\cal T}]\big)\varepsilon.\end{split}\ee where $\xslash=x^\mu\sigma_\mu$ and $\dslash=\sigma^\mu\partial_\mu$. Again, upon checking on-shell closure of Poincare SUSY transformations on canonical fields one reads \footnote{The equations of motion which are used are \be\begin{split} \square\phi^i &= - g\gamma^{ij}_{AB}[\bar\psi^A,\psi^B,\phi^j,{\cal T}] - g\gamma^i_{AB}[\bar\psi^A,\psi^B,\phi^8,{\cal T}] \cr - &g^2[\phi^j,\phi^k,[\phi^i,\phi^j,\phi^k,{\cal T}],{\cal T}] - g^2[\phi^j,\phi^8,[\phi^i,\phi^j,\phi^8,{\cal T}],{\cal T}], \cr \square\phi^8 &= - g\gamma^i_{AB}[\bar\psi^A,\psi^B,\phi^i,{\cal T}] - g^2[\phi^j,\phi^k,[\phi^8,\phi^j,\phi^k,{\cal T}],{\cal T}] \nonumber, \cr \sigma^\mu\partial_\mu\psi^A &= - g\gamma^{ij}_{AB}[\phi^i,\phi^j,\psi^B,{\cal T}] -g\gamma^i_{AB}[\phi^i,\phi^8,\psi^B,{\cal T}].\end{split}\nonumber \ee}
\be\begin{split} &[\delta_\epsilon,\delta_{\epsilon'}]\phi^i = \iota^\mu\partial_\mu\phi^i \cr &+ \Omega_{ij}(gt^{jklm}[\phi^k,\phi^l,\phi^m,{\cal T}] + t^{jkl}[\phi^k,\phi^l,\phi^8,{\cal T}])
\cr &+ \Omega_i(gt^{jkl}[\phi^j,\phi^k,\phi^l,{\cal T}])
,\end{split}\ee
\be\begin{split} &[\delta_\epsilon,\delta_{\epsilon'}]\phi^8 = \iota^\mu\partial_\mu\phi^8
\cr &+ \Omega_i(gt^{ijkl}[\phi^j,\phi^k,\phi^l,{\cal T}] + t^{ijk}[\phi^j,\phi^k,\phi^8,{\cal T}])
,\end{split}\ee
\be\begin{split} &[\delta_\epsilon,\delta_{\epsilon'}]\psi^A = \iota^\mu\partial_\mu\psi^A + (\Omega_{ij}\gamma^{ij}_{AB}+\Omega_i\gamma^i_{AB})\times \cr &\ \ \times(g\gamma^{kl}_{BC}[\phi^k,\phi^l,\psi^C,{\cal T}] + g\gamma^k_{BC}[\phi^k,\phi^8,\psi^C,{\cal T}]).\end{split}\ee
Forming algebra between Poincare SUSY and conformal SUSY transformations results as follows
\be\begin{split} &[\delta_\epsilon,\delta_{\varepsilon'}]\phi^i = Dx^\mu\partial_\mu\phi^i + \omega^{\mu\nu}(x_\mu\partial_\nu-x_\nu\partial_\mu)\phi^i\cr &+ \Omega_{ij}(\phi^i\!+g\xslash t^{jklm}[\phi^k,\phi^l,\phi^m,{\cal T}] + g\xslash t^{jkl}[\phi^k,\phi^l,\phi^8,{\cal T}])
\cr &+ \Omega_i(\phi^8+gt^{jkl}[\phi^j,\phi^k,\phi^l,{\cal T}])
,\end{split}\ee
\vspace*{-4mm}\be\begin{split} &[\delta_\epsilon,\delta_{\varepsilon'}]\phi^8 = Dx^\mu\partial_\mu\phi^8 + \omega^{\mu\nu}(x_\mu\partial_\nu-x_\nu\partial_\mu)\phi^8
\cr &+ \Omega_i(\phi^i+gt^{ijkl}[\phi^j,\phi^k,\phi^l,{\cal T}] + t^{ijk}[\phi^j,\phi^k,\phi^8,{\cal T}])
,\end{split}\ee
\vspace*{-4mm}\be\begin{split} &[\delta_\epsilon,\delta_{\varepsilon'}]\psi^A = Dx^\mu\partial_\mu\psi^A + \omega^{\mu\nu}(x_\mu\partial_\nu-x_\nu\partial_\mu+\epsilon_{\mu\nu\rho}\sigma^\rho)\psi^A \cr &\quad + (\Omega_{ij}\gamma^{ij}_{AB}+\Omega_i\gamma^i_{AB})\times\cr &\quad \times(\psi^B\!\!+g\xslash\gamma^{kl}_{BC}[\phi^k,\phi^l,\psi^C,{\cal T}]\! + \!g\xslash\gamma^k_{BC}[\phi^k,\phi^8,\psi^C,{\cal T}]).\end{split}\ee
Finally, forming algebra of conformal SUSY reads as
\be\begin{split} &[\delta_\varepsilon,\delta_{\varepsilon'}]\phi^i = \kappa^\mu(-2x_\mu x^\nu\partial_\nu+x^2\partial_\mu)\phi^i \cr &+ \Omega_{ij}(gx^2t^{jklm}[\phi^k,\phi^l,\phi^m,{\cal T}] + gx^2t^{jkl}[\phi^k,\phi^l,\phi^8,{\cal T}])
\cr &+ \Omega_i(x^2gt^{jkl}[\phi^j,\phi^k,\phi^l,{\cal T}]),\end{split}\ee
\vspace*{-4mm}\be\begin{split} &[\delta_\varepsilon,\delta_{\varepsilon'}]\phi^8 = \kappa^\mu(-2x_\mu x^\nu\partial_\nu+x^2\partial_\mu)\phi^8
\cr &+ \Omega_i(gx^2t^{ijkl}[\phi^j,\phi^k,\phi^l,{\cal T}] + gx^2t^{ijk}[\phi^j,\phi^k,\phi^8,{\cal T}])
,\end{split}\ee
\vspace*{-4mm}\be\begin{split} &[\delta_\varepsilon,\delta_{\varepsilon'}]\psi^A = \kappa^\mu(-2x_\mu x^\nu\partial_\nu+x^2\partial_\mu)\psi^A \cr &\qquad+  (\Omega_{ij}\gamma^{ij}_{AB}+\Omega_i\gamma^i_{AB})\times\cr &\qquad \times(gx^2\gamma^{kl}_{BC}[\phi^k,\phi^l,\psi^C,{\cal T}] + gx^2\gamma^k_{BC}[\phi^k,\phi^8,\psi^C,{\cal T}]).\end{split}\ee

The conserved charge of SUSY can be derived as
\be\begin{split} Q^A\!\! = \int {\rm d}^2x
{\rm Tr}\Big[\Big(\big(\dslash\phi^i &-igt^{ijkl}[\phi^j,\phi^k,\phi^l,{\cal T}] \cr &-igt^{ijk}[\phi^j,\phi^k,\phi^8,{\cal T}]\big)\gamma^i\cr + ( \dslash\phi^8 &-igt^{ijk}[\phi^i,\phi^j,\phi^k,{\cal T}])\Big)\times({\cal C}\psi^A)\Big],\end{split}\ee
\vspace*{-4mm}\be\begin{split} S^A\!\! =\! -\!\!\int {\rm d}^2x
{\rm Tr}\Big[\Big(\big(\xslash\dslash\phi^i-i\phi^i &-igt^{ijkl}[\phi^j,\phi^k,\phi^l,{\cal T}] \cr -i&gt^{ijk}[\phi^j,\phi^k,\phi^8,{\cal T}]\big)\gamma^i\cr + (\xslash\dslash\phi^8 -i \phi^8 -i &gt^{ijk}[\phi^i,\phi^j,\phi^k,{\cal T}])\Big)({\cal C}\psi^A)\Big].\end{split}\ee

As before, from ``$Q-Q$'' anti-commutator one reads conserved charge of translation $P^\mu=(H,P^m)$ \be\begin{split} H = \int{\rm d}^2x{\rm Tr}\Big[&\frac{1}{2}\pi^i\pi^i + \frac{1}{2}\pi^8\pi^8 + \frac{1}{2}\partial_i\phi^i\partial_i\phi^i + \frac{1}{2}\partial_i\phi^8\partial_i\phi^8 \cr - &i\psi^{\dagger A}\sigma^i\partial_i\psi^A \cr + \frac{1}{2} g\gamma^{ij}_{AB} \bar\psi^A&[\phi^i,\phi^j,\psi^B,{\cal T}] + \frac{1}{2} g\gamma^i_{AB} \bar\psi^A[\phi^i,\phi^8,\psi^B,{\cal T}] \cr + & \frac{1}{3!}g^2[\phi^i,\phi^j,\phi^k,{\cal T}]^2 + \frac{1}{3!}g^2[\phi^i,\phi^j,\phi^8,{\cal T}]^2 \Big],\end{split}\ee
\vspace*{-6mm}\be P_m = \int{\rm d}^2x{\rm Tr}\Big[\pi^i\partial_m\phi^i + \pi^8\partial_m\phi^8 + \psi^{\dagger A}\partial_m\psi^A\Big],\ee
and related algebraic extensions ${\cal Z}^{ij}_+$, ${\cal Z}^i_+$ and ${\cal Y}^{AB}_{\mu+}$.

From ``$Q-S$'' anti-commutator one reads conserved charges of dilatation $D$, spacetime rotation $M^{\mu\nu}$ as \bea D &=& \int{\rm d}^2x\big(x_\mu {\cal P}^\mu\big) , \cr M^{\mu\nu} &=& \int{\rm d}^2x\big(x^\mu{\cal P}^\nu-x^\nu {\cal P}^\mu),\eea where curly letters stands for density of conserved charges, {\it i.e.} taken off spatial integration. $R$-space conserved charge $J^{ij}$ and $J^i$ can be read as \bea J^{ij} &=& \int {\rm d}^2x{\rm Tr}\Big[\pi^{[i}\phi^{j]} + i\bar\psi^A\gamma^{ij}_{AB}\psi^B\Big],\\ J^{i\pm} &=& \pm\int{\rm d}^2x{\rm Tr}\Big[\pi^{[i}\phi^{8]} + i\bar\psi^A\gamma^i_{AB}\psi^B\Big].\eea
Algebraic extensions ${\cal Z}^{ij}_\mu$, ${\cal Z}^{i\pm}_\mu$, ${\cal Y}^{AB}$ and ${\cal Y}^{AB}_{\mu\nu}$ can also be derived from this anti-commutator.

Similarly, from ``$S-S$'' anti-commutator one reads conserved charges of SCT $K_\mu$ as \be K^\mu = \int{\rm d}^2x\big(-2x^\mu{\cal D} + x^2{\cal P}^\mu\big),\ee and related algebraic extensions ${\cal Z}^{ij}_-$, ${\cal Z}^i_-$ and ${\cal Y}^{AB}_{\mu-}$.

\subsubsection{$SO(4)\times SO(4)$ subgroup of $SO(8)$}
One proposes the following action which is manifestly invariant under $SO(1,2)\times SO(4)\times SO(4)$ \be\begin{split}\label{2-action} {\cal S} = \int{\rm d}^3x\ {\rm Tr}\Big[&\frac{1}{2}\partial^\mu\phi^i\partial_\mu\phi^i + \frac{1}{2}\partial^\mu\phi^{i'}\partial_\mu\phi^{i'} \cr +&
i\bar\psi^{\alpha\alpha'}\sigma^\mu\partial_\mu\psi^{\alpha\alpha'} + i\bar\psi^{\dot\alpha\dot\alpha'}\sigma^\mu\partial_\mu\psi^{\dot\alpha\dot\alpha'} \cr -& \frac{1}{2}\big(m\phi^i+g\epsilon^{ijkl}\phi^i[\phi^j\!,\phi^k\!,\phi^l\!,{\cal T}]\big)^2 \cr -& \frac{1}{2}\big(m\phi^{i'}+g\epsilon^{i'\!j'\!k'\!l'}\phi^{i'}[\phi^{j'}\!,\phi^{k'}\!,\phi^{l'}\!,{\cal T}]\big)^2 \cr - \bar\psi^{\alpha\alpha'}\!\!\big(m\psi^{\alpha\alpha'}\!\!\! +\! g\sigma^{ij}_{\alpha\beta}&[\phi^i,\phi^j,\psi^{\beta\alpha'}\!\!,{\cal T}]\!\! + \! g\sigma^{i'\!j'}_{\alpha'\beta'}[\phi^{i'}\!,\phi^{j'}\!,\psi^{\alpha\beta'}\!,{\cal T}] \big) \cr - \bar\psi^{\dot\alpha\dot\alpha'}\!\!\big(m\psi^{\dot\alpha\dot\alpha'}\!\!\! +\! g\sigma^{ij}_{\dot\alpha\dot\beta}&[\phi^i,\phi^j,\psi^{\dot\beta\dot\alpha'}\!\!,{\cal T}]\!\! + \! g\sigma^{i'\!j'}_{\dot\alpha'\dot\beta'}[\phi^{i'}\!,\phi^{j'}\!,\psi^{\dot\alpha\dot\beta'}\!,{\cal T}] \big) \cr -& [\phi^i,\phi^j,\phi^{k'}\!,{\cal
T}]^2 - [\phi^{i'}\!,\phi^{'j}\!,\phi^k,{\cal T}]^2\Big].\end{split}\ee
Scalars $\phi^i$ and $\phi^{i'}$ $(i=1,2,3,4)$ are in fundamental representation of each $SO(4)$ and $\epsilon^{ijkl}$ is its invariant tensor. Primed indices refers to second $SO(4)$. Spinors $\psi_{\tilde\alpha}^{\alpha\alpha'}$ and $\psi_{\tilde\alpha}^{\dot\alpha\dot\alpha'}$ are in $[(1/2,0),(1/2,0);\bf 2_r]$ and $[(0,1/2),(0,1/2);\bf 2_r]$ representation of $SU(2)\times SU(2)\times SU(2)\times SU(2)\times SL(2,\mathbb R)$. Upper undotted-dotted indices ($\alpha,\dot\alpha=1,2$) refer to Weyl and lower index ($\tilde\alpha=1,2$)refers to Majorana representation (for details of notation see \cite{TGMT,extensions,Sadri}).\footnote{Dimensional reduction of the above action on torus, has been proposed to be related to the DLCQ Hamiltonian of IIB string theory on the plane-wave background in sector with $N$ units of light-cone momenta as regularized D3-brane \cite{TGMT} and polarized non-BPS D0-branes \cite{DLCQ}.}

The following SUSY transformations leave the action invariant \bea\label{bosonic-Poincare-SUSY-2} \delta\phi^i &=& \bar\epsilon^{\alpha\dot\alpha'}\sigma^i_{\alpha\dot\beta}\psi^{\dot\beta\dot\alpha'}+ \bar\epsilon^{\dot\alpha\alpha'}\sigma^i_{\dot\alpha\beta}\psi^{\beta\alpha'},\\ \delta\phi^{i'} &=& \bar\epsilon^{\alpha\dot\alpha'}\sigma^{i'}_{\dot\alpha'\beta'}\psi^{\alpha\beta'}+ \bar\epsilon^{\dot\alpha\alpha'}\sigma^{i'}_{\alpha'\dot\beta'}\psi^{\dot\alpha\dot\beta'},\eea
\vspace*{-4mm}\be\begin{split} \delta\psi^{\alpha\alpha'} = \big(\dslash\phi^i + m\phi^i &+ g\epsilon^{ijkl}[\phi^j,\phi^k,\phi^l,{\cal T}]\big)\sigma^i_{\alpha\dot\beta}\epsilon^{\dot\beta\alpha'} \cr + \big(\dslash\phi^{i'}\!\! + m\phi^{i'}\!\! &+ g\epsilon^{i'\!j'\!k'\!l'}[\phi^{i'}\!,\phi^{j'}\!,\phi^{k'}\!,{\cal T}]\big)\sigma^{i'}_{\alpha'\dot\beta'}\epsilon^{\alpha\dot\beta'} \cr &+ [\phi^i,\phi^j,\phi^{i'},{\cal T}]\sigma^{ij}_{\alpha\beta}\sigma^{i'}_{\alpha'\dot\beta'}\epsilon^{\beta\dot\beta'} \cr &+[\phi^{i'}\!,\phi^{j'}\!,\phi^i\!,{\cal T}]\sigma^{i'\!j'}_{\alpha'\beta'}\sigma^i_{\alpha\dot\beta}\epsilon^{\dot\beta\beta'} ,\end{split}\ee \vspace*{-6mm}\be\begin{split}\delta\psi^{\dot\alpha\dot\alpha'} = \big(\dslash\phi^i &+ m\phi^i + g\epsilon^{ijkl}[\phi^j,\phi^k,\phi^l,{\cal T}]\big)\sigma^i_{\dot\alpha\beta}\epsilon^{\beta\dot\alpha'} \cr + \big(\dslash\phi^{i'}\!\! &+ m\phi^{i'}\!\! + g\epsilon^{i'\!j'\!k'\!l'}[\phi^{i'}\!,\phi^{j'}\!,\phi^{k'}\!,{\cal T}]\big)\sigma^{i'}_{\dot\alpha'\beta'}\epsilon^{\dot\alpha\beta'} \cr &+ [\phi^i,\phi^j,\phi^{i'},{\cal T}]\sigma^{ij}_{\dot\alpha\dot\beta}\sigma^{i'}_{\dot\alpha'\beta'}\epsilon^{\dot\beta\beta'} \cr &+[\phi^{i'}\!,\phi^{j'}\!,\phi^i\!,{\cal T}]\sigma^{i'\!j'}_{\dot\alpha'\dot\beta'}\sigma^i_{\dot\alpha\beta}\epsilon^{\beta\dot\beta'} ,\end{split}\ee
where parameters of SUSY transformations $\epsilon_\alpha^{\alpha\dot\alpha'}$ and $\epsilon_\alpha^{\dot\alpha\alpha'}$ are in $[(0,1/2),(1/2,0);\bf 2_r]$ and $[(1/2,0),(0,1/2);\bf 2_r]$ representations. With 16 different SUSY transformations, \eqref{2-action} is invariant under a large subgroup of $OSp(8|2)$.

However, if one turns off dimensionful parameter (the mass), then the action would be scale invariant which is believed to be invariant under full $SO(2,3)\times SO(4)\times SO(4)$ conformal transformation. Together with full conformal SUSY, it is invariant under a large subgroup of $OSp(8|4)$ SUSY group.

\section{Acknowledgement}
\begin{acknowledgments}
I am thankful to B. Acharya, E. Gava, C. Gowdigere, K. S. Narain, S. Randjbar-Daemi, S. Sheikh-Jabbari and H-U. Yee for insightful discussions and useful comments.

\end{acknowledgments}
\appendix
\section{A. The Construction of The Action}
An explicit formulation of an ${\cal N}=8$ SUSic action in 3D, realizing $SO(7)$ global symmetry as its $R$-symmetry, is presented here. Construction of the action with $SO(4)\times SO(4)$ $R$-symmetry is similar.

In order to write $SO(7)$ invariant non-linear interacting terms, one has to incorporate a non-trivial structure for fields. One assumes besides physical 3 dimensional spacetime there is also an internal 3-dimensional space, local coordinates of which are $\sigma^r (r=1,2,3)$. The canonical fields are then functions of both set of coordinates $\phi^i(x,\sigma)$, $\phi^8(x,\sigma)$ and $\psi^A_\alpha(x,\sigma)$. Now, the invariant tensors of physical space and internal space are $\eta^{\mu\nu},\epsilon^{\mu\nu\rho}$ and $\delta^{rs},\epsilon^{rst}$ respectively. Note that internal space derivations $\partial_r$ does not change the physical dimension.

The action is thus defined as \be {\cal S} = \int{\rm d}^3x\ {\cal L}\big[\varphi,\partial_\mu\varphi,\partial_r\varphi\big],\ee for $\varphi=\phi^i,\phi^8,\psi^A_\alpha$. The Lagrangian is constructed out of the canonical fields, their spacetime and internal derivatives and
their conjugate momenta glued by invariant tensors of demanded symmetry groups. One proposes the following interacting Lagrangian
\be\label{Classical-Lagrangian}\begin{split} {\cal L}\!\! &= \int{\rm
d}^3\sigma\Big(\frac{1}{2}\eta^{\mu\nu}\partial_\mu\phi^i\partial_\nu\phi^j\delta_{ij} + \frac{1}{2}\eta^{\mu\nu}\partial_\mu\phi^8\partial_\nu\phi^8 \cr
&\qquad\qquad + i\eta^{\mu\nu}\bar\psi^A_\alpha\sigma^\mu_{\alpha\beta}\partial_\nu\psi^A_\beta \cr -\frac{1}{2} & \big(m\phi^i\!\!+\!gt^{ijkl}\partial_r\phi^j\partial_s\phi^k\partial_t\phi^l\epsilon^{rst} \!\!+\!gt^{ijk}\partial_r\phi^j\partial_s\phi^k\partial_t\phi^8\epsilon^{rst}\big)^2
\cr -\frac{1}{2}& \big(m\phi^8\!\!+gt^{jkl}\partial_r\phi^j\partial_s\phi^k\partial_t\phi^l\epsilon^{rst}\big)^2 \cr -\frac{1}{2} &\bar\psi^A_\alpha\!\big(m\psi^A_\alpha \!\!+\! g\gamma^{ij}_{AB}\partial_r\phi^i\partial_s\phi^j\partial_t\psi^B_\alpha \!\!+\! g\gamma^i_{AB}\partial_r\phi^i\partial_s\phi^8\partial_t\psi^B_\alpha\big).\end{split}\ee
Noting appearance of Nambu 3-bracket defined as \cite{Nambu} \be
\epsilon^{mnp}\partial_m\varphi^I\partial_n\varphi^J\partial_p\varphi^K =
\big\{\varphi^I,\varphi^J,\varphi^K\big\},\ee which is skew-symmetric, satisfy Leibnitz derivation rule and a fundamental identity \cite{Takhtajan}, Lagrangian \eqref{Classical-Lagrangian} can be rewritten in a compact form
as \be\label{Classical-Lagrangian-2}\begin{split} {\cal L}\!\! &= \int{\rm
d}^3\sigma\Big(\frac{1}{2}\eta^{\mu\nu}\partial_\mu\phi^i\partial_\nu\phi^j\delta_{ij} + \frac{1}{2}\eta^{\mu\nu}\partial_\mu\phi^8\partial_\nu\phi^8 \cr
&\qquad\qquad\ + i\eta^{\mu\nu}\psi^{A\dagger}_\alpha({\cal C}\sigma_\mu)_{\alpha\beta}\partial_\nu\psi^A_\beta \cr & -\frac{1}{2}  \big(m\phi^i+gt^{ijkl}\{\phi^j,\phi^k,\phi^l\} +gt^{ijk}\{\phi^j,\phi^k,\phi^8\}\big)^2
\cr &-\frac{1}{2} \big(m\phi^8+gt^{jkl}\{\phi^j,\phi^k,\phi^l\}\big)^2 \cr -\frac{1}{2} &\bar\psi^A_\alpha\big(m\psi^A_\alpha + g\gamma^{ij}_{AB}\{\phi^i,\phi^j,\psi^B_\alpha\} + g\gamma^i_{AB}\{\phi^i,\phi^8,\psi^B_\alpha\}\big).\end{split}\ee

The Nambu brackets are linked to volume-preserving
diffeomorphisms \cite{Minic}.\footnote{Volume-preserving diffeomorphisms, part of full diffeomorphisms, on internal space are described by a differentiable map $\sigma^m\rightarrow f^m(\sigma)$, such that $\{f^1,f^2,f^3\}=1$. Transformations involve two independent functions, ${\cal G}_1$ and ${\cal G}_2$. The generators are \be G = \epsilon^{mnp}\,\partial_m{\cal G}_1\, \partial_n {\cal G}_2\partial_p = G^p\partial_p, \ee satisfying $\partial_pG^p=0$. An arbitrary scalar function $\Phi(\sigma^m)$ infinitesimally transforms as \be\label{continuous-transformation} \delta_\xi\Phi = G\cdot\Phi = \big\{{\cal
G}_1,{\cal G}_2,\Phi\big\}.\ee} Thus, the bracket form of the action
\eqref{Classical-Lagrangian-2} implies that it could be invariant under such an infinite-dimensional group of transformations in internal space. In fact if one supposes internal space is mapped into a Euclidean target space whose local coordinates are $\phi^I$, then the triple product $\{\phi^I,\phi^J,\phi^K\}$ is also invariant under volume-preserving diffeomorphisms. Actually due to fundamental identity it transforms as a scalar \cite{Minic}. Similar argument works for Grassmann coordinates $\psi^A$. Hence, such diffeomorphisms do not change the bracket forms and thus leave the action invariant. Therefore, the bracket form of the action is the reminiscent of a symmetry in internal space. The group theory and algebra of the generators of volume-preserving diffeomorphisms is elaborated to some extend \cite{Minic}. It suggests that there is a new kind of symmetry based on a new composition rule whose algebra is given by triple commutator.

However, it is difficult to quantize the theory, because of the non-linearity
of the equations of motion and difficulty in solving them. On the other hand, in this construction there are infinite number of degrees of freedom. In fact, there is a regularization procedure for theories which have 2-dimensional area preserving diffeomorphisms where the surface has a compact topology \cite{Hoppe}. They can be regularized by applying the Goldstone-Hoppe map between representation theories of the algebra of area-preserving diffeomorphisms and the $N\rightarrow\infty$ limit of Lie algebras. It then  instructs one to perform the following prescription; functions are mapped to finite sized matrices, Poisson brackets get replaced by matrix commutators and surface integration by trace over matrix indices \cite{dWHN,Taylor}. After regularizing the classical theory the resulting theory is a system which has a finite number of degrees of freedom.

One wishes to apply similar procedure for volume-preserving diffeomorphisms of a 3-dimensional manifold \cite{TGMT}. For simplicity one supposes it has the
topology of a 3-sphere. In this case the manifold can be described by a unit
sphere with an $SO(4)$ invariant canonical Nambu form. Functions on this
manifold can be described in terms of functions of 4 Cartesian coordinates
$\zeta^i$ on the unit 3-sphere satisfying
\bea\label{classical-1nd-property} \delta^{ij}\zeta_i\zeta_j &=& 1, \\
\label{classical-2nd-property}\{\zeta^i,\zeta^j,\zeta^k\} &=&
\epsilon^{ijkl}\zeta^l.\eea
Furthermore, by the definition of Nambu bracket, it is possible to introduce a fixed
function, $\zeta^5$, in such a way that 3-bracket promotes to a 4-bracket which
is nicer practically. It is called odd-to-even embedding
\cite{Curtright-Zachos} \be\label{odd-even-embedding}
\big\{\zeta^i,\zeta^j,\zeta^k\big\}\rightarrow
\big\{\zeta^i,\zeta^j,\zeta^k,\zeta^5\big\}.\ee The Nambu 4-bracket can now be
resolved to a fully anti-symmetrized sum of strings of Poisson brackets.

Indeed considering \eqref{classical-1nd-property},
\eqref{classical-2nd-property} and \eqref{odd-even-embedding} it resembles the same algebraic structure as that defined by 4-commutator of gamma matrices of
$Spin(4)$ \cite{Guralnik,half-BPS}. In fact they satisfy \bea
\delta^{ij}\gamma_i\gamma_j &=& 4, \\\ [\gamma^i,\gamma^j,\gamma^k,\gamma^5]
&=& -4!\epsilon^{ijkl}\gamma^l.\eea It is instructive to associate coordinates on $S^3$ with the gamma matrices by this correspondence \bea
\zeta^i&\rightarrow&\Upsilon^i,\\ \zeta^5&\rightarrow&\Upsilon^5,\eea where
$\Upsilon$ are generalized gamma matrices of $SO(4)$ in $N$ dimensional
representation. These matrices define a non-associative algebra
as totally anti-symmetric trilinear product (alternatively called 3-algebra) \cite{half-BPS,Tanzini} \be [\ \ ,\ ,\ ,{\cal T}]:{\cal
A}\times{\cal A}\times{\cal A}\rightarrow{\cal A},\ee where ${\cal
T}=\Upsilon^5$ is a definite fixed matrix [appendix B].

Thus, generally any function $\varphi$ on this manifold can be expanded as a sum of spherical harmonics of $SO(4)$ \be \varphi(\zeta^i)=c^{lm_1m_2}Y_{lm_1m_2}(\zeta^i).\ee The spherical
harmonics can in turn be written as sum of monomials in the coordinate
functions \be Y_{lm_1m_2}(\zeta^i)= t_{lm_1m_2}^{i_1\dots
i_l}\zeta_{i_1}\dots\zeta_{i_l},\ee where the coefficients $t$ are symmetric
and traceless. Thus the matrix approximations to each of the spherical
harmonics with can be constructed through \bea Y_{lm_1m_2}(\zeta^i) \rightarrow
\left({\bf Y}_{lm_1m_2}\right)_{rs} = {\bf t}_{lm_1m_2}^{i_1\dots
i_{l}}\Big(\Upsilon_{i_1}\dots\Upsilon_{i_{l}}\Big)_{rs}.\ \eea The matrix
approximation of the spherical harmonics can be used to construct matrix
approximations to an arbitrary function as \be \varphi(\zeta)\rightarrow\Phi\,_{rs} = {\bf
c}^{lm_1m_2}({\bf Y}_{lm_1m_2})_{rs}.\ee

It instructs one to perform the following prescription to regularize the theory \cite{TGMT}; \\
Differentiable functions are mapped to matrices \be\label{matrix-fields} \varphi(x,\zeta) \rightarrow \Phi(x)_{rs} = {\bf c}_{lm_1m_2}(x){\bf t}^{lm_1m_2}_{i_1i_2\cdots i_l}\Big(\Upsilon^{i_1}\Upsilon^{i_2}\cdots\Upsilon^{i_l}\Big)_{rs},\ee where $i=1,2,3,4$, $|m_1,m_2|\leq l$, $l\leq n$ and matrix indices $r,s=1,\cdots,N$ where $N$ is the size of matrices as a function of $n$ to be determined momentarily. The number of independent degrees of freedom is counted by ${\bf c}(x)$. Nambu bracket is promoted to matrix 4-commutators \be
\big\{\phi^I,\phi^J,\phi^K\big\} \rightarrow N[\phi^I,\phi^J,\phi^K,{\cal
T}].\ee Fixed matrix ${\cal T}$ is
introduced in the recipe of regularization along the definition of
4-commutator, to have a well-behaved one \cite{TGMT}. It has to be precisely
defined. \\ The volume integration is replaced by trace over matrices \be\int{\rm d}^3\sigma\rightarrow\frac{1}{N}{\rm Tr}.\ee Performing all this one is led to \eqref{massive-action}. It defines a field theory with finite number of degrees of freedom. In principle the quantization of such a theory is straightforward.

The fields are now Hermitian $N\times N$ matrices. Naively, theory has a symmetry group $U(N)$ of global (internal) transformations. From the solution to the equations of motion, it becomes clear that ${\cal T}$ is a traceless matrix which squares to identity matrix. Indeed it is proportional to generalized $Spin(4)$ chirality matrix. One can use $U(N)$ rotations to brings ${\cal T}^5$ in this form. Then, upon precisely defining it, this group is spontaneously broken to ${\cal S}_N$, the group of permutations of $N$ objects \cite{half-BPS}. The Goldstone bosons are just entries of the matrix fields.

\section{B. The Non-Associative Algebra}\label{appendixB}
The 4-commutator is defined as fully anti-symmetric product of four objects
\be\label{4-commutator} [{\cal O}_1,{\cal O}_2,{\cal O}_3,{\cal O}_4] =
\epsilon^{ijkl}{\cal O}_i\cdot{\cal O}_j\cdot{\cal O}_k\cdot{\cal O}_l.\ee In
the matrix realization of the 4-commutator, ${\cal O}$'s are ordinary square
matrices and $\cdot$ is ordinary associative matrix multiplication. It
satisfies generalized Jacobi identity \be \epsilon^{ijklmnp}[[{\cal O}_i,{\cal
O}_j,{\cal O}_k,{\cal O}_l],{\cal O}_m,{\cal O}_n,{\cal O}_p]=0,\ee and by-part
integration. It forfeits Leibnitz derivation property, {\it i.e.} the
associativity is compromised \cite{TGMT}. The trace of a 4-commutator is zero.
It has a proper classical limit and there is a $c$-number for the 4-commutator,
{\it i.e.} there is an identity element such that $[\ ,\ ,\ ,{\bf 1}]=0$.
Definition \eqref{4-commutator} can be written as \be\label{4-commutator-alter}
[{\cal O}_1,\!{\cal O}_2,\!{\cal O}_3,\!{\cal O}_4]\!\!\sim\!
\epsilon^{ijkl}[{\cal O}_i,\!{\cal O}_j,\!{\cal O}_k,\!{\cal O}_l] \!\!\sim\!
\epsilon^{ijkl}[{\cal O}_i,\!{\cal O}_j][{\cal O}_k,\!{\cal O}_l].\ee

In order that 4-commutator defines an algebraic structure one supposes that it yields a fifth element \be [{\cal O}_1,{\cal O}_2,{\cal O}_3,{\cal O}_4] =
{\cal O}_5.\ee Using \eqref{4-commutator-alter}, it can be rewritten as \be
[{\cal O}_i,{\cal O}_j,{\cal O}_k,{\cal O}_l] = \epsilon_{ijkl}{\cal O}_5.\ee
The aim now is to look for a set of five independent elements satisfying the
above non-associative algebraic structure and close into each other. Through
representation theory of $Spin(n)$, noting the fact that multiplication of gamma matrices is not associative and the fact that $\epsilon^{ijkl}$ is an invariant tensor of $Spin(4)$, the idea is to identify ${\cal O}$'s with four gamma matrices $Spin(4)$.

To see this more systematically, one starts with 5 dimensional Euclidean space. Its isometry group is $SO(5)$ and it has 4 dimensional spinor representation $r$ on which $Spin(5)$ gamma matrices $\gamma^m$ act. One defines generalized gamma matrices $\Gamma^m$ of the algebra ${\cal A}^5$ of $N'\times N'$ hermitian matrices as $n$ fold direct tensor product of gamma matrices and identity matrix \be \Gamma^m = \left(\gamma_m\otimes{\bf
1}\otimes\dots\otimes{\bf 1} + \dots + {\bf 1}\otimes\dots\otimes{\bf
1}\otimes\gamma_m\right)_{sym},\ee act on the symmetrized $n$-fold tensor
product of smallest irreducible spinor representation $r$, $\left(r^{\otimes n}\right)_{sym}$. The dimension of representation is
$N'=(n+1)(n+2)(n+3)/6$ \cite{Ramgoolam}. This matrices have the following
properties \cite{half-BPS} \bea \delta_{mn}\Gamma^m\Gamma^n &=& n(n+4){\bf
1_{_{N'}}}, \\\ [\Gamma^m,\Gamma^n,\Gamma^p,\Gamma^q] &=&
8(n+2)!!\epsilon^{mnpqr}\Gamma^r,\eea where $\delta^{mn}$ and
$\epsilon^{mnpqr}$ are invariant tensors of $SO(5)$.

Now, one singles out one direction to reduce isometries to $SO(4)$, say $5th$
direction. Under chirality projection ${\cal P}_\pm$, $r$ decomposes into $r_+$
and $r_-$, positive and negative chirality 2-dimensional spinor representations
of $Spin(4)$, where four gamma matrices $\gamma^i$ and chirality matrix
$\gamma^5$ act. Most generally $N\times N$ Hermitian gamma matrices
$\Upsilon$'s, defined as projected $n$-fold direct tensor product of gamma
matrices and identity matrix  \cite{Ramgoolam,Berman-Copland}\bea \Upsilon_i &=& {\cal P}_{\cal R}\Gamma^i{\cal P}_{\cal R}, \\
\Upsilon_5 &=& {\cal P}_{\cal R}\Gamma_5{\cal P}_{\cal R}. \eea  In fact they
are in $N\times N$ dimensional either irreducible or reducible representations
of $Spin(4)$ and chirality matrix. They are in the matrix algebra ${\cal A}^4$.
The projection operator ${\cal P}_{\cal R}={\cal P}_{{\cal R}_+}+{\cal
P}_{{\cal R}_-}$ \be {\cal P}_{{\cal R}_\pm} = \left({\cal P}_+^{\otimes(n\pm
1)/2}\otimes{\cal P}_-^{\otimes(n\mp 1)/2}\right),\ee makes sure that the
generalized gamma matrices act within symmetrized tensor product space of
reducible spinor representation ${\cal R}={\cal R}_++{\cal R}_-$ defined as \be
{\cal R} \!=\! \left(r_+^{\otimes(n+1)/2}\otimes
r_-^{\otimes(n+1)/2}\right)_s\!\! \oplus \left(r_-^{\otimes(n-1)/2}\otimes
r_+^{\otimes(n-1)/2}\right)_s,\ee where $r_+$ and $r_-$ are irreducible
positive and negative chirality 2-dimensional spinor representation of
$Spin(4)$ and ${\cal R}_+$ and ${\cal R}_-$ are irreducible representation with
$(2j_L,2j_R)$ respectively $(n+1/2,n-1/2)$ and $(n-1/2,n+1/2)$. The dimension
of ${\cal R}$
is $N=(n+1)(n+3)/2$. It can be shown \cite{half-BPS,Basu-Harvey,Guralnik} \bea \Upsilon^5\Upsilon^5 &=& {\bf 1_{_N}}, \\
\delta^{ij}\Upsilon^i\,\Upsilon^j &=& N\cdot{\bf 1_{_N}}.\eea Furthermore
$\Upsilon^5$ and $\Upsilon^i$ anti-commute and also fulfill
\bea\label{non-associative-structure}
[\Upsilon^i,\Upsilon^j,\Upsilon^k,\Upsilon^5] &=& -\epsilon^{ijkl}\Upsilon^l,
\\\ (or\ [\Upsilon^i,\Upsilon^j,\Upsilon^k,\Upsilon^l\,] &=&
+\epsilon^{ijkl}\Upsilon^5.)\qquad\nonumber \eea
Thus, the 4-commutator can be thought of as acts like $[\,\ ,\ ,\
,\Upsilon^5]:{\cal A}\times{\cal A}\times{\cal A}\rightarrow{\cal A}$ and drop
projection.

One also demands this algebra reproduces the classical algebra of
differentiable functions in the large $N$ limit. Thus the projections should be imposed properly to single out proper operators \cite{Ramgoolam}. The number of operators surviving in $End({\cal R^\pm})$ is $n(n+1)(n+2)/6$ and the number of them in $Hom({\cal R}^\pm,{\cal R}^\mp)$ is $(n+1)(n+2)(n+3)/6$. Furthermore, the matrices should act in the same manner on ${\cal R}^+$ and ${\cal R}^-$. Thus, one sums each state of $End({\cal R^+})$ with the corresponding one from $End({\cal R^-})$ and so on, giving the total number of degrees of freedom as $(n+1)(n+2)(2n+3)/6$. However, the algebra does not close under multiplication, and one has to project back into ${\cal A}_4$ after multiplication. This product is in general non-associative.

Finally, the number of degrees of freedom of Hermitian $N\times N$ matrix fields which are spanned in gamma basis \eqref{matrix-fields} and obey the
non-associative algebraic structure \eqref{non-associative-structure}, is the number of independent coefficients ${\bf c}_{lm_1m_2}(x)$ of \eqref{matrix-fields} which can be calculated as
\be \sum_{l=0}^{n}(2l+1)^2 \sim n^3 \sim N^{3/2}.\ee
It implies number of degrees of freedom scales like $N^{3/2}$ not $N^2$ (for similar result see also \cite{Berman-Copland}).


\end{document}